\documentclass{article}
\usepackage{amssymb}

\usepackage{amsmath}
\usepackage{doublespace}

\input{tcilatex}

\begin{document}

\bigskip

\bigskip

\bigskip

\bigskip

\bigskip

\bigskip

\begin{center}
\bigskip \textbf{\ New exact Taylor's expansions and simple solutions to PDEs%
}

\bigskip

\bigskip Moawia Alghalith
\end{center}

\bigskip

\textbf{Abstract}. We provide new \textit{exact} Taylor's series with fixed
coefficients and without the remainder. We demonstrate the usefulness of
this contribution by using it to obtain very simple solutions to
(non-linear) PDEs. We also apply the method to the portfolio model.

\bigskip


AMS \# 41A58

\bigskip

\bigskip

\bigskip

\pagebreak

\section{\protect\bigskip {\protect\Large Introduction}}

In this paper, we provide an \textit{exact} Taylor's expansion ( Taylor
(1715)) with \textit{constant} coefficients and without the remainder. In
doing so, we provide an explicit (closed) form of the remainder. Needless to
say, this pioneering contribution is extremely useful in many applications,
such as the areas of optimization, integration, and partial differential
equations PDEs, since it transforms any arbitrary function to an exact
explicit quasi-linear form. Consequently, this method will simplify the
solutions to cumbersome integrals, PDEs and optimization problems (see
Alghalith (2008) for potential applications). Consequently, this
contribution can be applied to many areas in operations research, such as
optimization, stochastic models, and statistics.

We apply our approach to the heat equation and the
reaction-convection-diffusion equation. In doing so, we use it to transform
a (non-linear) PDE to a non-differential equation or an ordinary
differential equation ODE. We also apply our method to the dynamic portfolio
model in finance. In so doing, unlike previous literature, we derive an
explicit solution to the investor's optimal portfolio when the utility
function is unknown.

\section{\protect\Large \protect\bigskip The new expansions}

\textbf{Theorem}: A sufficiently differentiable function with a compact
support $f\left( .\right) $ is given by

$(i)$ $f\left( x\right) =a_{1}+a_{2}x+a_{2}\left( x+a_{3}\right) \ln \left(
x+a_{3}\right) ,$

$(i)$ $f\left( x,y\right) =a_{1}+a_{2}x+a_{3}y+a_{2}\left( x+a_{3}\right)
\ln \left( x+a_{3}\right) $, where $a$ is a constant.

\textbf{Proof.}

$(i)$ Consider these Taylor's expansions

\begin{equation}
f\left( x\right) =f\left( c\right) +f^{\prime }\left( c\right) \left(
x-c\right) +R_{1}\left( x\right) ,c\neq 0,  \label{1}
\end{equation}%
\begin{equation}
f\left( x\right) =f\left( c\right) +R_{2}\left( x\right) ,
\end{equation}%
where $R$ is the remainder and $c$ is a constant. Taking the derivatives of
the remainders w.r.t. $x$ yields%
\begin{equation*}
R_{1}^{\prime }\left( x\right) =f^{\prime }\left( x\right) -f^{\prime
}\left( c\right) ,
\end{equation*}

\begin{equation*}
R_{2}^{\prime }\left( x\right) =f^{\prime }\left( x\right) .
\end{equation*}%
Thus%
\begin{equation}
R_{2}^{\prime }\left( x\right) -R_{1}^{\prime }\left( x\right) =f^{\prime
}\left( c\right) .  \label{100}
\end{equation}%
Dividing both sides of $\left( \ref{100}\right) $ by $x-u+\alpha $ and using
the mean value theorem, we obtain 
\begin{equation*}
R_{1}^{\prime \prime }\left( x\right) =\frac{R_{2}^{\prime }\left( x\right)
-R_{1}^{\prime }\left( x\right) }{x-u+\alpha }=\frac{f^{\prime }\left(
c\right) }{x-u+\alpha },c\leq u\leq x,x-u+\alpha \neq 0.
\end{equation*}%
where $\alpha $ is a positive constant. Integrating the above equation yields%
\begin{eqnarray}
R_{1}\left( x\right) &=&\int\limits_{c}^{x}\int\limits_{c}^{x}\frac{%
f^{\prime }\left( c\right) }{x-u+\alpha }dudu=  \notag \\
&&f^{\prime }\left( c\right) \left[ \alpha \ln \left( \alpha \right)
+x-\left( \left( x-c+\alpha \right) \ln \left( x-c+\alpha \right) +c\right) %
\right] .  \label{3a}
\end{eqnarray}%
Substituting $\left( \ref{3a}\right) $ into $\left( \ref{1}\right) ,$ we
obtain

\begin{eqnarray*}
f\left( x\right) &=&f\left( c\right) +f^{\prime }\left( c\right) \left(
x-c\right) + \\
&&f^{\prime }\left( c\right) \left[ \alpha \ln \left( \alpha \right)
+x-\left( \left( x-c+\alpha \right) \ln \left( x-c+\alpha \right) +c\right) %
\right] .
\end{eqnarray*}%
We can rewrite the above equation as 
\begin{equation}
f\left( x\right) =a_{1}+a_{2}x+a_{2}\left( x+a_{3}\right) \ln \left(
x+a_{3}\right) ,  \label{4}
\end{equation}%
where $a$ is a constant.$\square $

$\left( ii\right) $ As before we consider these Taylor's expansions 
\begin{equation}
f\left( x,y\right) =f\left( c_{1},c_{2}\right) +f_{x}\left(
c_{1},c_{2}\right) \left( x-c_{1}\right) +f_{y}\left( c_{1},c_{2}\right)
\left( y-c_{2}\right) +R_{1}\left( x,y\right) ,  \label{9}
\end{equation}

\begin{equation}
f\left( x,y\right) =f\left( c_{1},c_{2}\right) +R_{2}\left( x,y\right) .
\end{equation}%
Taking the partial derivatives of the remainders w.r.t. $x$ yields

\begin{equation*}
R_{1x}\left( x,y\right) =f_{x}\left( x,y\right) -f_{x}\left(
c_{1},c_{2}\right) ,
\end{equation*}

\begin{equation*}
R_{2x}\left( x,y\right) =f_{x}\left( x,y\right) .
\end{equation*}%
Therefore%
\begin{equation*}
R_{2x}\left( x,y\right) -R_{1x}\left( x,y\right) =f_{x}\left(
c_{1},c_{2}\right) .
\end{equation*}%
Thus%
\begin{equation*}
R_{1xx}\left( .\right) =\frac{f_{x}\left( c_{1},c_{2}\right) }{x-u+\alpha }.
\end{equation*}%
Integrating yields

\begin{equation*}
R_{1}\left( x,y\right) =\int\limits_{c_{1}}^{x}\int\limits_{c_{1}}^{x}\frac{%
f_{x}\left( c_{1},c_{2}\right) }{x-u+\alpha }dxdx=
\end{equation*}

\begin{equation}
f_{x}\left( c_{1},c_{2}\right) \left[ \alpha \ln \left( \alpha \right)
+x-\left( \left( x-c_{1}+\alpha \right) \ln \left( x-c_{1}+\alpha \right)
+c_{1}\right) \right] .  \label{5}
\end{equation}%
Substituting $\left( \ref{5}\right) $ into $\left( \ref{9}\right) $ yields%
\begin{equation}
f\left( x,y\right) =a_{1}+a_{2}x+a_{3}y+a_{2}\left( x+a_{3}\right) \ln
\left( x+a_{3}\right) .\square  \label{50}
\end{equation}%
The extension to a multiple-variable function is straightforward.

\section{\protect\Large Practical examples}

In this section, we demonstrate the revolutionary nature of this
contribution. In particular, we use it to transform sophisticated PDEs to
non-differential equations or first-order ODEs.

\textbf{The reaction-convection-diffusion equation:}

The reaction-convection-diffusion equation has many applications in
operations research, physics and finance. An example of such an equation is%
\begin{equation}
V_{t}+rxV_{x}+\frac{1}{2}\sigma ^{2}x^{2}V_{xx}-rV=0,  \label{20}
\end{equation}%
subject to a boundary condition. Using $\left( \ref{50}\right) $ yields%
\begin{equation}
V\left( x,t\right) =a_{1}+a_{2}x+a_{3}t+a_{2}\left( x+a_{3}\right) \ln
\left( x+a_{3}\right) .  \label{51}
\end{equation}%
Taking the partial derivatives of $\left( \ref{51}\right) $ yields%
\begin{equation}
V_{x}\left( x,t\right) =a_{2}\left[ 2+\ln \left( x+a_{3}\right) \right] ,
\label{21}
\end{equation}%
\begin{equation}
V_{t}\left( x,t\right) =a_{3},  \label{22}
\end{equation}

\begin{equation}
V_{xx}\left( x,t\right) =\frac{a_{2}}{x+a_{3}}.  \label{23}
\end{equation}%
Substituting $\left( \ref{21}\right) -\left( \ref{23}\right) $ into $\left( %
\ref{20}\right) $, we obtain%
\begin{equation*}
a_{3}+rxa_{2}\left[ 2+\ln \left( x+a_{3}\right) \right] +\frac{1}{2}\frac{%
a_{2}\sigma ^{2}x^{2}}{x+a_{3}}-rV=0.
\end{equation*}%
Needless to say, we transformed $\left( \ref{20}\right) $ to a
non-differential equation, such that $V$ is given explicitly.

\textbf{The heat equation:}

An example of a diffusion equation is the heat equation%
\begin{equation}
V_{t}-kV_{xx}=0.  \label{26}
\end{equation}%
As before, substituting $\left( \ref{23}\right) $ into $\left( \ref{26}%
\right) $ yields%
\begin{equation*}
V_{t}-\frac{ka_{2}}{x+a_{3}}=0.
\end{equation*}%
Clearly, we transformed the heat equation to an ODE. Moreover, $V$ is
continuously differentiable and thus, a unique classical solution exists.
Also, this method is easily applicable to non-linear PDEs.

\textbf{The portfolio model:}

We first provide a brief description of the baseline portfolio model (for an
excellent review, see Kolm et al (2014)). Our approach may also be
applicable to other variants of this model (see, for example, Palczewski et
al (2015) and Alghalith (2012), among others). The risk-free asset price
process is given by $S_{0}=e^{\int\limits_{t}^{T}r_{s}ds},$ where $r_{t}\in
C_{b}^{2}\left( R\right) $ is the risk-free rate of return. The dynamics of
the risky asset price are given by

\begin{equation}
dS_{s}=S_{s}\left( \mu _{s}ds+\sigma _{s}dW_{s}\right) ,
\end{equation}%
where $\mu _{s}\in C_{b}^{2}\left( R\right) $ and $\sigma _{s}\in
C_{b}^{2}\left( R\right) $ are the rate of return and the volatility,
respectively; $W_{s}$ is a Brownian motion on the probability space $\left(
\Omega ,\mathcal{F},\mathcal{F}_{s},P\right) ,$ where $\left\{ \mathcal{F}%
_{s}\right\} _{t\leq s\leq T}$ is the augmentation of filtration.

The wealth process is given by

\begin{equation}
X_{T}^{\pi }=x+\int\limits_{t}^{T}\left\{ rX_{s}^{\pi }+\left( \mu
_{s}-r_{s}\right) \pi _{s}\right\} ds+\int\limits_{t}^{T}\pi _{s}\sigma
_{s}dW_{s},
\end{equation}%
where $x$ is the initial wealth, $\left\{ \pi _{s},\mathcal{F}_{s}\right\}
_{t\leq s\leq T}$ is the portfolio process, with $E\int\limits_{t}^{T}\pi
_{s}^{2}ds<\infty .$ The trading strategy $\pi _{s}\in \mathcal{A}\left(
x,y\right) $ is admissible$.$

The investor maximizes the expected utility of the terminal wealth

\begin{equation*}
V\left( t,x\right) =\underset{\pi }{Sup}E\left[ U\left( X_{T}^{\pi }\right)
\mid \mathcal{F}_{t}\right] ,
\end{equation*}%
where $V\left( .\right) $ is the value function, $U\left( .\right) $ is a
continuous, bounded and strictly concave utility function. Under well-known
assumptions, the value function satisfies the Hamilton-Jacobi-Bellman PDE%
\begin{equation*}
V_{t}+rxV_{x}+\pi _{t}^{\ast }\left( \mu _{t}-r_{t}\right) V_{x}+
\end{equation*}

\begin{equation*}
\frac{1}{2}\pi _{t}^{\ast 2}\sigma _{t}^{2}V_{xx}=0;V\left( T,x\right)
=U\left( x\right) .
\end{equation*}%
Following the procedure in the previous examples, we can rewrite the above
Hamilton-Jacobi-Bellman PDE as%
\begin{equation*}
V_{t}+a_{2}\left[ rx+\pi _{t}^{\ast }\left( \mu _{t}-r_{t}\right) \right] %
\left[ 2+\ln \left( x+a_{3}\right) \right] +\frac{1}{2}\frac{a_{2}\pi
_{t}^{\ast 2}\sigma _{t}^{2}}{x+a_{3}}=0,
\end{equation*}%
which is clearly a first-order ODE. The optimal portfolio is given by%
\begin{equation*}
\pi _{t}^{\ast }=-\frac{\left( \mu _{t}-r_{t}\right) V_{x}}{\sigma
_{t}^{2}V_{xx}}=-\frac{a_{2}\left( \mu _{t}-r_{t}\right) \left[ 2+\ln \left(
x+a_{3}\right) \right] }{\sigma _{t}^{2}\left( \frac{a_{2}}{x+a_{3}}\right) }%
.
\end{equation*}%
Thus, we explicitly expressed the optimal portfolio as a function of the
initial wealth, even if the utility function is unknown.

\end{document}